\documentclass[aps,prx,onecolumn,showpacs,amsmath,notitlepage,amssymb]{revtex4-1}

\usepackage{bm}        
\usepackage{amssymb}   
\usepackage{amsmath}
\usepackage{amsthm}
\usepackage{graphicx}  
\usepackage{color}
\usepackage{colortbl}
\usepackage{ascmac}

\usepackage[T1]{fontenc}

\definecolor{dartmouthgreen}{rgb}{0.05, 0.5, 0.06} 

\newcommand{\hh}{\mathcal{H}}
\newcommand{\vv}{\mathcal{V}}
\newcommand{\cd}{\mathbb{C}^d}
\newcommand{\cdim}[1]{\mathbb{C}^{#1}}
\newcommand{\symd}[1]{\bigvee_{#1}\mathbb{C}^d}

\newcommand{\sep}[2]{\mathbf{SEP}\left(#1:#2\right)}
\newcommand{\linop}[1]{\mathbf{L}\left(#1\right)}
\newcommand{\herm}[1]{\mathbf{Herm}\left(#1\right)}
\newcommand{\pos}[1]{\mathbf{Pos}\left(#1\right)}
\newcommand{\dop}[1]{\mathbf{S}\left(#1\right)}
\newcommand{\puredop}[1]{\mathbf{P}\left(#1\right)}

\newcommand{\PhiS}[1]{\Phi_{#1}}

\newcommand{\ketPhiS}[1]{\ket{\Phi}_{#1}}

\newcommand{\phiS}[1]{\phi_{#1}}

\newcommand{\phiSI}[2]{\phi_{#1}^{(#2)}}

\newcommand{\psiS}[1]{\psi_{#1}}
\newcommand{\psiSI}[2]{\psi_{#1}^{(#2)}}

\newcommand{\idop}{\mathbb{I}}

\newcommand{\tr}[1]{tr\left[#1\right]}
\newcommand{\ptr}[2]{tr_{#1}\left[#2\right]}
\newcommand{\conv}[1]{{\rm conv}\left(#1\right)}

\newcommand{\lpnorm}[2]{\left\|#2\right\|_{#1}}

\newcommand{\fidelity}[2]{F\left(#1,#2\right)}

\newcommand{\bra}[1]{\langle {#1} |}
\newcommand{\ket}[1]{| {#1} \rangle}
\newcommand{\braket}[2]{\langle {#1} |{#2} \rangle}
\newcommand{\ketbra}[1]{| {#1} \rangle\langle {#1} |}

\newtheorem{lemma}{Lemma}

\newtheorem{theorem}{Theorem}

\begin{document}

\title{On the hardness of conversion from entangled proof into separable one}
\author{Seiseki Akibue}
\email{seiseki.akibue@ntt.com}
 \affiliation{NTT Communication Science Laboratories, NTT Corporation\\3-1, Morinosato-Wakamiya, Atsugi, Kanagawa 243-0198, JAPAN}
 
\author{Go Kato}%
\email{go.kato@nict.go.jp}
\affiliation{%
 Advanced ICT Research Institute, NICT\\
  4--2--1, Nukui-Kitamachi, Koganei, Tokyo 184-8795, Japan
}
 
 \author{Seiichiro Tani}
\email{seiichiro.tani@ntt.com}
 \affiliation{NTT Communication Science Laboratories, NTT Corporation\\3-1, Morinosato-Wakamiya, Atsugi, Kanagawa 243-0198, JAPAN}

\date{\today}

\begin{abstract}
A quantum channel whose image approximates the set of separable states is called a disentangler, which plays a prominent role in the investigation of variants of the computational model called Quantum Merlin Arthur games, and has potential applications in classical and quantum algorithms for the separability testing and NP-complete problems. So far, two types of a disentangler, constructed based on $\epsilon$-nets and the quantum de Finetti theorem, have been known; however, both of them require an exponentially large input system. Moreover, in 2008, John Watrous conjectured that any disentangler requires an exponentially large input system, called the disentangler conjecture.
In this paper, we show that both of the two known disentanglers can be regarded as examples of a {\it strong disentangler}, which is a disentangler approximately breaking entanglement between one output system and the composite system of another output system and the arbitrarily large environment. Note that the strong disentangler is essentially an approximately entanglement-{\it breaking} channel while the original disentangler is an approximately entanglement-{\it annihilating} channel, and the set of strong disentanglers is a subset of disentanglers. As a main result, we show that the disentangler conjecture is true for this subset, the set of strong disentanglers, for a wide range of approximation parameters without any computational hardness assumptions.
\end{abstract}

\maketitle
\section{Introduction}
Entanglement is an essential resource that provides non-classical phenomena in quantum mechanics and advantages in quantum information processing over classical one. Thus, testing whether a given quantum state is entangled or separable is a fundamental task for investigating and utilizing quantum nature. One of the primitive ways to detect entanglement is using an entanglement witness \cite{HHH96}. However, it is known that the number of entanglement witnesses represented by positive maps necessary for detecting any (even robustly) entangled state in $\cdim{d}\otimes\cdim{d}$ is $\exp(\Omega(d^3/\log d))$~\cite{AS17}. Moreover, if we formalize the quantum separability testing as a promise problem via the weak membership problem within an inverse polynomial precision, it has been shown to be NP-hard \cite{G04, G10}.
On the other hand, such complex structures of separable states provide benefits to the computation when we use a separable state as {\it quantum proof} in the computational model called Quantum Merlin Arthur games (QMA) \cite{KMY01, ABDFS08}. Indeed, proof encoded in a log-size separable state is sufficient for solving NP-complete problems, 3-COL \cite{BT09} and 3-SAT \cite{B08,GNN12}, whereas it seems impossible to solve such NP-complete problems with proof encoded in a log-size entangled state \cite{MW05}.

\begin{figure}
 \centering
  \includegraphics[height=.17\textheight]{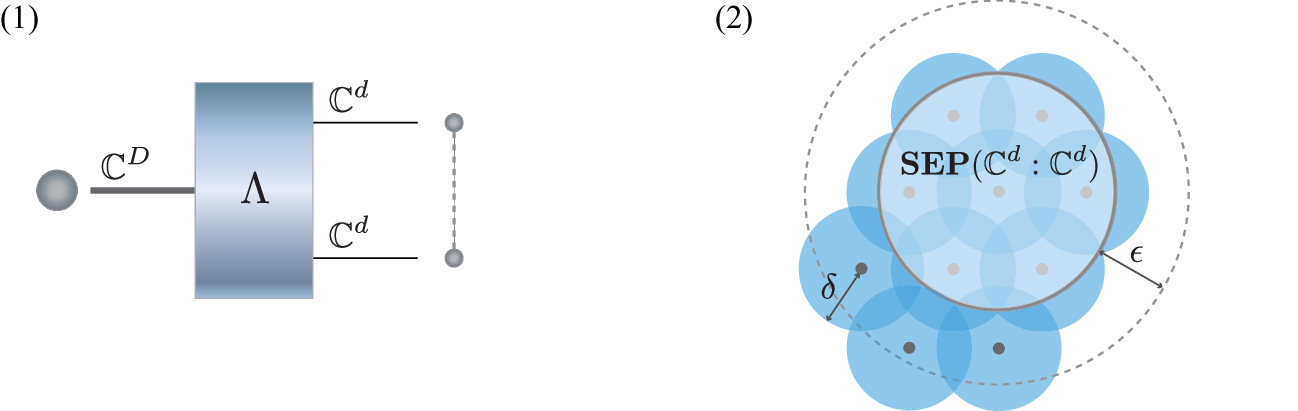}
  \caption{Graphical representations of an $(\epsilon,\delta)$-disentangler $\Lambda$. (1) $\Lambda$ produces only approximately separable states. (2) The points and small disks surrounding them represent producible states by $\Lambda$ and their $\delta$-neighborhoods, respectively. The large disk surrounded by the solid circle and that surrounded by the dashed circle represent the set of separable states $\sep{\cdim{d}}{\cdim{d}}$ and its $\epsilon$-neighborhood, respectively. The first condition of the disentangler requires that all the producible states by $\Lambda$ reside in the  $\epsilon$-neighborhood of $\sep{\cdim{d}}{\cdim{d}}$. The second condition requires that $\delta$-neighborhoods of the producible states cover $\sep{\cdim{d}}{\cdim{d}}$. More precise definitions are given in Section \ref{section:disentangler}. }
\label{fig:disent}
\end{figure} 

The remarkable computational power provided by proof encoded in a separable state has induced the {\it disentangler conjecture}, which states the difficulty of converting the set of entangled states into that of separable states \cite{ABDFS08}. More precisely, the conjecture states that exponentially large input dimension $D$, i.e., $\log D=\Omega(d)$ with respect to dimension $d$ of one output system, is necessary for realizing the quantum channel called an $(\epsilon,\delta)$-disentangler, whose output state is an almost separable state within precision $\epsilon$ and approximates an arbitrary separable state within precision $\delta$ as shown in Fig.\ref{fig:disent}. Thus, the disentangler is an approximated-entanglement-{\it annihilating} channel, whose exact version is defined in \cite{MZ10}, in contrast to the well known entanglement-{\it breaking} channel \cite{H99, HSR03}. 
Despite its simplicity and importance, the conjecture is far from a complete proof or a falsification. Indeed, there exist only a few known ways to construct a disentangler including a construction based on $\epsilon$-nets and that based on the quantum de Finetti theorem \cite{ABDFS08}. Moreover, the only nonexistence proofs without assuming any computational hardness assumption are given by \cite{ABDFS08} for the nonexistence of $(0,0)$-disentanglers on a finite dimensional Hilbert space and by \cite{HNW18} for the nonexistence of $(\epsilon,\delta)$-disentanglers with $\epsilon+\delta=O\left(\frac{1}{d^2}\right)$ having a polynomial input dimension, i.e., it requires $\log D=\Omega\left(\frac{(\log d)^2}{poly\log\log d}\right)$.

\begin{figure}
 \centering
  \includegraphics[height=.13\textheight]{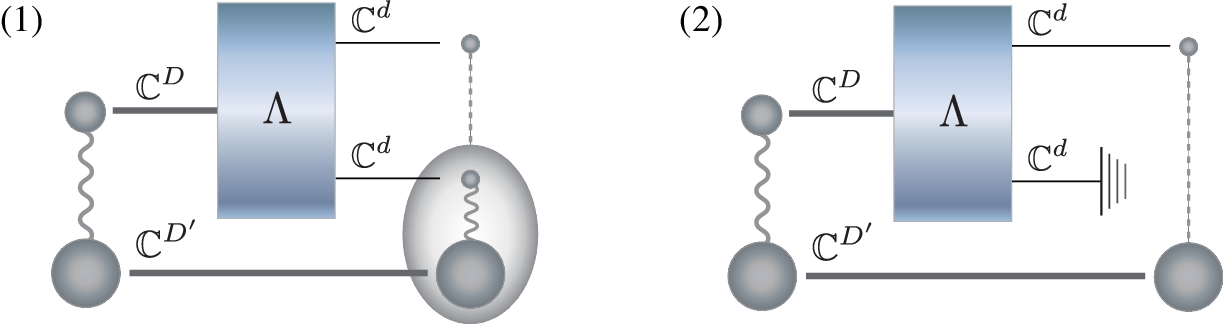}
  \caption{Graphical representations of strong disentangler $\Lambda$. (1) $\Lambda$ satisfies not only the conditions of the disentangler but also the condition such that $\Lambda$ approximately breaks entanglement between one output system and the composite system of another output system and the arbitrarily large environment. (2) If we discard one output system of $\Lambda$, it is an approximately entanglement-breaking channel. More precise definitions are given in Section \ref{section:disentangler}.} 
\label{fig:sdisent}
\end{figure} 

In this paper, we define the quantum channel we call an $(\epsilon,\delta)$-{\it strong disentangler}, which is an $(\epsilon,\delta)$-disentangler and breaks entanglement between one output system and the composite system of another output system and the arbitrarily large environment within precision $\epsilon$ as shown in Fig.\ref{fig:sdisent}. Then, we show that the two known disentanglers are examples of the strong disentangler. Note that the strong disentangler is essentially an approximately entanglement-{\it breaking} channel while the original disentangler is an approximately entanglement-{\it annihilating} channel, and the set of strong disentanglers is a subset of disentanglers. As a main result, we show, without any computational hardness assumption, that the disentangler conjecture is true for the strong disentangler. More precisely, we obtain the following theorem.
\begin{screen}
\textbf{Theorem 1.} (informal version)
{\it For any constants $\epsilon,\delta\geq0$ such that $\epsilon+\sqrt{\delta}<1$, every $(\epsilon,\delta)$-strong disentangler satisfies $\log D=\Omega(d)$ as $d\rightarrow\infty$.}
\end{screen}

\section{Notations and reviews}
\label{section:notation}
In this section, we summarize basic notations used throughout the paper and review the quantum de Finetti theorem, which is deeply related to the disentangler. Remark that we consider only finite dimensional Hilbert spaces.

\subsection{Basic notations}
$\linop{\hh}$, $\herm{\hh}$ and $\pos{\hh}$ represent the set of linear operators, hermitian operators and positive semidefinite operators on Hilbert space $\hh$, respectively. The set of quantum states is represented by that of density operators, defined as
\begin{equation}
 \dop{\hh}:=\left\{\rho\in\pos{\hh}:\tr{\rho}=1\right\}.
\end{equation}
The set of pure quantum states is represented by
\begin{equation}
 \puredop{\hh}:=\left\{\rho\in\dop{\hh}:\tr{\rho^2}=1\right\}.
\end{equation}
It is known that $\dop{\hh}$ and $\puredop{\hh}$ are compact and convex. Note that the compactness of subsets in $\linop{\hh}$ and the (uniform) continuity of functions on subsets in $\linop{\hh}$ (or their product space) are defined with respect to the standard topology on $\linop{\hh}$, i.e., the topology induced by a norm on $\linop{\hh}$.
Sometimes, a pure state is alternatively represented by complex unit vector $\ket{\phi}\in\hh$ such that $\braket{\phi}{\phi}=1$. Unnormalized complex vector is denoted with tilde, e.g., $\ket{\tilde{\eta}}$. For pure state $\ket{\phi}$ (or complex vector $\ket{\tilde{\eta}}$), we denote its density operator (or rank-$1$ linear operator) as $\phi:=\ket{\phi}\bra{\phi}\in\puredop{\hh}$ (or $\tilde{\eta}:=\ket{\tilde{\eta}}\bra{\tilde{\eta}}$.) 
We sometimes denote a subscript to emphasize the system where a state resides, e.g. $\rho_A$ implies $\rho_A\in\dop{\hh_A}$.
A reduced density operator of pure state $\ket{\Phi}_{AB}\in\hh_A\otimes\hh_B$ is denoted as
\begin{equation}
 \Phi_A:=\ptr{B}\Phi,
\end{equation}
where the partial trace $tr_B:\linop{\hh_A\otimes\hh_B}\rightarrow\linop{\hh_A}$ is defined by $tr_B=id_A\otimes tr$ by using the identity map $id:\linop{\hh_A}\rightarrow \linop{\hh_A}$ and the trace map $tr:\linop{\hh_B}\rightarrow\mathbb{C}$. We use a subscript to identify the tracing system and to emphasize the system where the linear map acts.

The set of separable states is denoted by
\begin{eqnarray}
 \sep{\hh_A}{\hh_B}&:=& \conv{\left\{\phiS{A}\otimes\psiS{B}:\phiS{A}\in\puredop{\hh_A},\psiS{B}\in\puredop{\hh_B}\right\}}.
\end{eqnarray}
It is known that $\sep{\hh_A}{\hh_B}$ is compact and convex.

Any physical process can be represented by a quantum channel, defined as a linear completely positive and trace-preserving (CPTP) map $\Gamma:\linop{\hh_{A}}\rightarrow\linop{\hh_B}$, where the initial state of the process is regarded as input state $\rho\in\dop{\hh_A}$ of quantum channel $\Gamma$ and the final state of the process is given by output state $\Gamma(\rho)\in\dop{\hh_B}$.
For any linear map $\Gamma:\linop{\hh_{A}}\rightarrow\linop{\hh_B}$, we can define corresponding Choi operator $J(\Gamma)\in\linop{\hh_A\otimes\hh_B}$ such as
\begin{equation}
 J(\Gamma):=\sum_{i,j}\ket{i}\bra{j}_A\otimes\Gamma(\ket{i}\bra{j}_{A}),
\end{equation}
where $\{\ket{i}\}_i$ is an orthonormal basis of $\hh_A$.

We often measure the distinguishability of quantum states by using a norm on $\linop{\hh}$ called the trace distance, defined by
\begin{equation}
 \lpnorm{tr}{M}:=\frac{1}{2}\lpnorm{1}{M},
\end{equation}
where $\lpnorm{1}{M}:=\tr{\sqrt{MM^\dag}}$. Note that for any states $\rho,\sigma\in\dop{\hh}$, it holds that $\lpnorm{tr}{\rho-\sigma}\leq1$, and the equality holds if and only if $\rho$ and $\sigma$ are perfectly distinguishable. 

We also use the fidelity function to measure the distinguishability, defined by
\begin{equation}
 \fidelity{\rho}{\sigma}:=\max|\braket{\Phi^{\rho}}{\Phi^\sigma}|^2
\end{equation}
for any quantum states $\rho,\sigma\in\dop{\hh_A}$, where pure states $\ket{\Phi^{\rho}}_{AB}$ and $\ket{\Phi^{\sigma}}_{AB}$ on $\hh_A\otimes\hh_B$ represent purifications of $\rho$ and $\sigma$ respectively, i.e., $\Phi^\rho_A=\rho$ and $\Phi^\sigma_A=\sigma$, and the maximization is taken over all the purifications.
The trace distance and the fidelity of two states $\rho,\sigma\in\dop{\hh}$ are related as follows:
\begin{equation}
\label{eq:FG}
 1-\sqrt{\fidelity{\rho}{\sigma}}\leq\lpnorm{tr}{\rho-\sigma}\leq\sqrt{1-\fidelity{\rho}{\sigma}},
\end{equation}
where the right equality holds when $\rho$ and $\sigma$ are pure states.

Both the trace distance and the fidelity function satisfy the {\it monotone} property that any physical process cannot increase the distinguishability of quantum states as follows: for any two states $\rho,\sigma\in\dop{\hh_A}$ and any quantum channel $\Gamma:\linop{\hh_{A}}\rightarrow\linop{\hh_B}$,
\begin{equation}
 \lpnorm{tr}{\rho-\sigma}\geq\lpnorm{tr}{\Gamma(\rho)-\Gamma(\sigma)}\ \wedge\ \fidelity{\rho}{\sigma}\leq\fidelity{\Gamma(\rho)}{\Gamma(\sigma)}.
\end{equation}

\subsection{Entanglement-breaking channel}
$\Lambda:\dop{\hh_E}\rightarrow\dop{\hh_B}$ is called an entanglement-breaking channel if $\Lambda$ is a linear CPTP map and $(\Lambda\otimes id_{E'})(\rho)\in\sep{B}{E'}$ for all input states $\rho\in\dop{\hh_E\otimes\hh_{E'}}$. The equivalent condition can be written as $\frac{1}{\dim\hh_E}J(\Lambda)\in\sep{E}{B}$ by using the Choi operator $J(\Lambda)$ of $\Lambda$. Define the set of Choi operators of entanglement-breaking channels by $\mathcal{E}_{E\rightarrow B}$, which is equivalent to
\begin{equation}
 \mathcal{E}_{E\rightarrow B}=\left\{(\dim\hh_E)\sigma:\sigma\in\sep{E}{B}\wedge \ptr{B}{\sigma}=\frac{1}{\dim\hh_E}\idop_{E}\right\}.
\end{equation}
By definition, it is clear that $\mathcal{E}_{E\rightarrow B}$ is convex. It is also obvious that $\mathcal{E}_{E\rightarrow B}$ is compact since it is the intersection of a compact set and a closed set. Owing to the Caratheodory's theorem, we can represent the set as
\begin{equation}
\mathcal{E}_{E\rightarrow B}=\left\{\sum_{i=1}^s\psiSI{B}{i}\otimes \tilde{\eta}_i:\{\tilde{\eta}_i\}_i\ {\rm is\ rank-}1{\rm \ POVM}\right\}, 
\end{equation}
with $s\leq(\dim\hh_B\dim\hh_E)^2$.

\subsection{Quantum de Finetti theorem}
Quantum de Finetti theorem is obtained by extending the (classical) de Finetti theorem and asserts that a reduced state of any quantum state $\rho$ on symmetric subspace $\symd{n}\subseteq(\cd)^{\otimes n}$ is approximately a probability mixture of independent and identically distributed (i.i.d.) states $\int d\mu(\phi)\phi^{\otimes k}$ for some probability measure $\mu$ \cite{CKMR07, C10}, i.e.,
\begin{equation}
\label{eq:deFinetti}
 \lpnorm{tr}{\ptr{[n-k]}{\rho}-\int d\mu(\phi)\phi^{\otimes k}}<\frac{kd}{n},
\end{equation}
where $[n]=\{1,2,\cdots,n\}$ and $tr_{[n-k]}$ represents the partial trace of the first $(n-k)$ system.
The theorem not only provides a basis for ``information-based interpretations" of quantum mechanics \cite{CFS02} but also has several applications to information processing tasks where an i.i.d. state is favorable, including quantum key distribution \cite{R07}, quantum tomography \cite{R07} and quantum hypothesis testing \cite{BP10}. 

Note that in Eq.\eqref{eq:deFinetti}, we refer an improved bound given in \cite{C10} comparing to the original bound given in \cite{CKMR07}. Furthermore, in \cite[Theorem 3]{C10}, the quantum de Finetti theorem has been generalized to the following form: for any finite-dimensional Hilbert space $\hh$ and any quantum state $\rho$ on $\symd{n}\otimes \hh$, there exists some probability measure $\mu$ and quantum state $\sigma:\puredop{\cd}\rightarrow\dop{\hh}$ such that
\begin{equation}
 \label{eq:generaldeFinetti}
 \lpnorm{tr}{\ptr{[n-k]}{\rho}-\int d\mu(\phi)\phi^{\otimes k}\otimes\sigma(\phi)}<\frac{kd}{n}.
\end{equation}
Remark that Eq.\eqref{eq:generaldeFinetti} implies Eq.\eqref{eq:deFinetti} when $\hh=\mathbb{C}$. This generalized theorem tells us how well $\ptr{[n-k]}{\rho}$ can be approximated by a (not necessarily i.i.d.) separable state, which ensures the completeness of the separability testing based on the $k$-extendibility \cite{APF04}. In the next section, we construct a disentangler based on these quantum de Finetti theorems.

\subsection{$\epsilon$-net}
The $\epsilon$-net is a subset of a set which can approximate any element of the set within precision $\epsilon$ with respect to some distance. In this paper, we use $\epsilon$-net $I\subseteq\puredop{\cd}$ of $\puredop{\cd}$ with respect to the trace distance. That is, $I$ satisfies
\begin{equation}
 \forall\phi\in\puredop{\cd},\exists \hat{\phi}\in I,\lpnorm{tr}{\hat{\phi}-\phi}\leq\epsilon.
\end{equation}
The minimum size $|I|$ of the $\epsilon$-net was given by us as follows:

\begin{lemma}
 \cite[Lemma 5]{SGS23-2}
\label{lemma:net}
 For any $\epsilon\in(0,1]$ and an integer $d\geq2$ specified below, the minimum size $|I|$ of the $\epsilon$-net of $\puredop{\cd}$ is bounded by
\begin{equation}
 2(d-1)\log_2\left(\frac{1}{\epsilon}\right)\leq\log_2|I|\leq  2(d-1)\log_2\left(\frac{1}{\epsilon}\right)+\log_2(5d\ln d).
\end{equation} 
\end{lemma}

\section{Disentangler and strong disentangler}
\label{section:disentangler}
In this section, we review the definition of the disentangler and give two explicit constructions of it. We also define a strong disentangler and verify that the two disentanglers are examples of the strong disentangler.
\subsection{Definitions}
Linear CPTP map $\Lambda:\dop{\cdim{D}}\rightarrow\dop{\cd\otimes \cd}$ is called an $(\epsilon,\delta)$-disentangler if it satisfies the following two conditions:
\begin{screen}
\begin{enumerate}
 \item $\forall \rho\in\dop{\cdim{D}},\min_{\sigma\in\sep{\cd}{\cd}}\lpnorm{tr}{\Lambda(\rho)-\sigma}\leq\epsilon$
 \item $\forall \sigma\in\sep{\cd}{\mathbb{C}^d}, \min_{\rho\in\dop{\cdim{D}}}\lpnorm{tr}{\Lambda(\rho)-\sigma}\leq\delta$.
\end{enumerate}
\end{screen}

Linear CPTP map $\Lambda:\dop{\cdim{D}}\rightarrow\dop{\cd\otimes \cd}$ is called an $(\epsilon,\delta)$-strong disentangler if it satisfies the following two conditions:
\begin{screen}
\begin{enumerate}
 \item $\forall \hh_R$ s.t. $\dim{\hh_R}<\infty, \forall \rho\in\dop{\cdim{D}\otimes\hh_R},\min_{\sigma\in\sep{\cd}{\cd\otimes\hh_R}}\lpnorm{tr}{(\Lambda\otimes id_{R})(\rho)-\sigma}\leq\epsilon$
 \item $\forall \sigma\in\sep{\cd}{\mathbb{C}^d}, \min_{\rho\in\dop{\cdim{D}}}\lpnorm{tr}{\Lambda(\rho)-\sigma}\leq\delta$.
\end{enumerate}
\end{screen}
By definition, an $(\epsilon,\delta)$-strong disentangler is an $(\epsilon,\delta)$-disentangler. 

The disentangler conjecture is the following conjecture: 
\begin{screen}
\begin{description}
 \item[Disentangler conjecture \cite{ABDFS08}] For any constants $\epsilon,\delta\geq0$ such that $\epsilon+\delta<1$, every $(\epsilon,\delta)$-disentangler satisfies $\log D=\Omega(d)$ as $d\rightarrow\infty$.
\end{description}
\end{screen}

\subsection{Constructions}
We review two types of a disentangler, constructed based on $\epsilon$-nets and the quantum de Finetti theorem.
which were suggested in \cite{ABDFS08}. We provide their explicit construction with bounds for the input dimension.
By slightly modifying the proof, we can verify that these two disentanglers are also strong disentanglers with the same approximation parameters $\epsilon$ and $\delta$.

\subsubsection{Disentangler based on $\epsilon$-net}
Let $I$ and $\{\ket{e_{\hat{\phi}}}\in\cdim{|I|}\}_{\hat{\phi}\in I}$ be a $\sqrt{\delta}$-net of $\puredop{\cd}$ and an orthonormal basis, respectively. Let $D=|I|d$. Define
\begin{equation}
 \Lambda(\rho):=\sum_{\hat{\phi}\in I}\hat{\phi}\otimes\ptr{1}{\ketbra{e_{\hat{\phi}}}\otimes\idop_d\rho}.
\end{equation}
Since the image of $\Lambda$ contains only separable states, $\Lambda$ satisfies the first condition of the disentangler with $\epsilon=0$.
We can verify the second condition as follows:
Let $\sigma=\sum_jp(j)\phi^{(j)}\otimes\psi^{(j)}$. Since the convex hull of a $\sqrt{\delta}$-net forms a $\delta$-net of $\puredop{\cd}$ \cite[Theorem 1]{SGS23-2}, we can find probability distribution $q(\hat{\phi}|j)$ such that $\lpnorm{tr}{\phi^{(j)}-\sum_{\hat{\phi}\in I}q(\hat{\phi}|j)\hat{\phi}}\leq\delta$. By letting the input state be $\rho=\sum_jp(j)\sum_{\hat{\phi}\in I}q(\hat{\phi}|j)\ketbra{e_{\hat{\phi}}}\otimes\psi^{(j)}$, we can verify that
\begin{equation}
 \lpnorm{tr}{\Lambda(\rho)-\sigma}\leq\sum_jp(j)\lpnorm{tr}{\sum_{\hat{\phi}\in I}q(\hat{\phi}|j)\hat{\phi}\otimes\psi^{(j)}-\phi^{(j)}\otimes\psi^{(j)} }=
 \sum_jp(j)\lpnorm{tr}{\sum_{\hat{\phi}\in I}q(\hat{\phi}|j)\hat{\phi}-\phi^{(j)} }\leq\delta,
\end{equation}
where we use the triangle inequality in the first inequality.

By using Lemma \ref{lemma:net}, this construction provides a $(0,\delta)$-disentangler with $\log_2D\leq(d-1)\log_2\left(\frac{1}{\delta}\right)+\log_2(5d\ln d)+\log_2 d$.

\subsubsection{Disentangler based on  quantum de Finetti theorem}
Let $U:\cdim{\dim(\symd{n})}\rightarrow\symd{n}$ be an isometry operator and $D=\dim(\symd{n})d$. Define
\begin{equation}
 \Lambda(\rho):=\ptr{[n-1]}{(U\otimes\idop_d)\rho(U^\dag\otimes\idop_d)}.
\end{equation}
Since the image of $\Lambda$ contains any separable states, $\Lambda$ satisfies the second condition with $\delta=0$.
We can verify the first condition as follows: for any $\rho$, by letting $\rho'=(U\otimes\idop_d)\rho(U^\dag\otimes\idop_d)\in\dop{\symd{n}\otimes\cd}$ and applying Eq.~\eqref{eq:generaldeFinetti}, we obtain
\begin{equation}
\min_{\sigma\in\sep{\cd}{\cd}}\lpnorm{tr}{\Lambda(\rho)-\sigma}=\min_{\sigma\in\sep{\cd}{\cd}}\lpnorm{tr}{\ptr{[n-1]}{\rho'}-\sigma}
<\frac{d}{n}.
\end{equation}
Thus, the second condition is satisfied with $\epsilon=\frac{d}{n}$. This construction provides 
a $(\epsilon,0)$-disentangler with $\log_2D<(d-1)\log_2\left(e(1+\frac{d}{d-1}\left(\frac{1}{\epsilon}+\frac{1}{d}\right)\right)+\log_2 d$, where we use an inequality
$
\begin{pmatrix}
 n\\k
\end{pmatrix}
<\left(\frac{n e}{k}\right)^k.
$

\section{Proof of the conjecture for strong disentanglers}
In this section, we prove the following main theorem.
\begin{theorem}
 For any constants $\epsilon,\delta\geq0$ such that $\epsilon+\sqrt{\delta}<1$, every $(\epsilon,\delta)$-strong disentangler satisfies $\log _2D\geq\frac{d-1}{2}\log_2\left(\frac{1}{\Delta}\right)-2\log_2d$ as $d\rightarrow\infty$, where $\Delta=1-\left(1-\epsilon-\sqrt{\delta}\right)^2$.
\end{theorem}
\begin{proof}
 
Define induced CPTP map $\Gamma:\dop{\cdim{D}}\rightarrow\dop{\cd}$ by $\Gamma(\rho)=\ptr{2}{\Lambda(\rho)}$ with an $(\epsilon,\delta)$-strong disentangler $\Lambda:\dop{\cdim{D}}\rightarrow\dop{\cd\otimes\cd}$. Then the conditions of the strong disentangler imply that
\begin{enumerate}
 \item $\forall \hh_R$ s.t. $\dim\hh_R<\infty, \forall \rho\in\dop{\cdim{D}\otimes\hh_R},\min_{\sigma\in\sep{\cd}{\hh_R}}\lpnorm{tr}{(\Gamma\otimes id_R)(\rho)-\sigma}\leq\epsilon$,
 \item $\max_{\phi\in\puredop{\cd}} \min_{\rho\in\dop{\cdim{D}}}\lpnorm{tr}{\Gamma(\rho)-\phi}\leq\delta$.
\end{enumerate}
Since $\{\Gamma(\rho):\rho\in\dop{\cdim{D}}\}$ is a compact convex subset of $\dop{\cd}$, by using a lemma about the convex approximation \cite[
Lemma 3]{SGS23-2}, the second condition is equivalent to
\begin{eqnarray}
 \max_{\phi,\psi\in\puredop{\cd}}\left(\tr{\phi\psi}-\max_{\rho\in\dop{\cdim{D}}}\tr{\psi\Gamma(\rho)}\right)\leq\delta\\
 \label{eq:cond2}
 \Leftrightarrow\min_{\psi\in\puredop{\cd}}\max_{\rho\in\dop{\cdim{D}}}\fidelity{\psi}{\Gamma(\rho)}\geq1-\delta.
\end{eqnarray}

Let $V:\hh_A\rightarrow\hh_B\otimes\hh_E$ with $\dim\hh_A=D$, $\dim\hh_B=d$, and $\dim\hh_E=Dd$ be an Stinespring dillation of $\Gamma$, i.e.,  $\Gamma(\rho)=\ptr{E}{V\rho V^\dag}$. Let $\vv$ be the range of $V$. Eq.~\eqref{eq:cond2} implies that
\begin{equation}
\label{eq:cond2-1}
 \forall \ket{\phi}\in\hh_B, \exists \ket{\psi}\in\hh_E,\exists \ket{\Phi}\in\vv, \braket{\phi\psi}{\Phi}\geq \sqrt{1-\delta}.
\end{equation}
Let $\{\phi^{(i)}\in\puredop{\hh_B}\}_i$, $\{\psi^{(i)}\in\puredop{\hh_E}\}_i$ and $\{\Phi^{(i)}\in\puredop{\vv}\}_i$  be an $\epsilon'$-net of $\puredop{\hh_B}$ and the corresponding states satisfying Ineq.~\eqref{eq:cond2-1}.
Let $\rho=\Psi\in\puredop{\hh_A\otimes\hh_R}$ with $(V\otimes\idop_R)\ket{\Psi}=\sum_{i}\sqrt{p(i)}\ket{\Phi^{(i)}}_{BE}\ket{i}_R$ be an input state of $\Gamma$. Then, the first condition of the strong disentangler implies that
\begin{equation}
\label{eq:cond1-1}
\max_p \min_{\sigma\in\sep{\hh_B}{\hh_R}}\lpnorm{tr}{\ptr{E}{(V\otimes\idop_R)\Psi (V^\dag\otimes\idop_R)}-\sigma}\leq\epsilon.
\end{equation}
Since $\bra{\hat{\Psi}}(V\otimes\idop_R)\ket{\Psi}\geq\sqrt{1-\delta}$ with $\ket{\hat{\Psi}}=\sum_{i}\sqrt{p(i)}\ket{\phi^{(i)}}_B\ket{\psi^{(i)}}_E\ket{i}_R$, Eq.~\eqref{eq:FG} implies that 
\begin{equation}
\label{eq:cond0}
\lpnorm{tr}{(V\otimes\idop_R)\Psi (V^\dag\otimes\idop_R)-\hat{\Psi}} \leq\sqrt{\delta}.
\end{equation}
By using the triangle inequality and the monotonicity of the trace distance, Eq.~\eqref{eq:cond1-1} and Eq.~\eqref{eq:cond0} imply that
\begin{equation}
\max_p \min_{\sigma\in\sep{\hh_B}{\hh_R}}\lpnorm{tr}{\ptr{E}{\hat{\Psi}}-\sigma}\leq\epsilon+\sqrt{\delta}.
\end{equation}
From now on, we assume that $\epsilon+\sqrt{\delta}<1$.
By using Eq.~\eqref{eq:FG}, we obtain
\begin{equation}
\min_p \max_{\sigma\in\sep{\hh_B}{\hh_R}}\fidelity{\ptr{E}{\hat{\Psi}}}{\sigma}\geq\left(1-\epsilon-\sqrt{\delta}\right)^2.
\end{equation}

By using Lemma \ref{lemma:fidelity} shown in Appendix, we obtain
\begin{eqnarray}
&&\min_p \max_{J(\mathcal{E})\in\mathcal{E}_{E\rightarrow B}}\tr{J(\mathcal{E})\ptr{R}{\hat{\Psi}}}\geq\left(1-\epsilon-\sqrt{\delta}\right)^2\\
&\Leftrightarrow&\min_p \max_{J(\mathcal{E})\in\mathcal{E}_{E\rightarrow B}}\sum_ip(i)\tr{J(\mathcal{E})\phi^{(i)}_B\otimes\psi^{(i)}_E}\geq\left(1-\epsilon-\sqrt{\delta}\right)^2.
\end{eqnarray}
Since the domain of $p$ and $J(\mathcal{E})$ is convex and compact sets and the target function is bilinear, we can apply the minimax theorem.
After applying the theorem, we obtain
\begin{equation}
 \max_{J(\mathcal{E})\in\mathcal{E}_{E\rightarrow B}}\min_i\tr{J(\mathcal{E})\phi^{(i)}_B\otimes\psi^{(i)}_E}\geq\left(1-\epsilon-\sqrt{\delta}\right)^2.
\end{equation}
By taking the limit of $\epsilon'\rightarrow0$, we obtain
\begin{eqnarray}
&& \max_{J(\mathcal{E})\in\mathcal{E}_{E\rightarrow B}}\min_{\phi\in\puredop{\hh_B}}\max_{\psi\in\puredop{\hh_E}}\tr{J(\mathcal{E})\phi_B\otimes\psi_E}\geq\left(1-\epsilon-\sqrt{\delta}\right)^2\\
 &\Leftrightarrow& \max_{J(\mathcal{E})\in\mathcal{E}_{E\rightarrow B}}\min_{\phi\in\puredop{\hh_B}}\max_{\psi\in\puredop{\hh_E}}\tr{\mathcal{E}(\psi)\phi}\geq\left(1-\epsilon-\sqrt{\delta}\right)^2.\\
\label{eq:cond1+2}
 &\Leftrightarrow& \max_{J(\mathcal{E})\in\mathcal{E}_{E\rightarrow B}}\min_{\phi\in\puredop{\hh_B}}\max_{\psi\in\puredop{\hh_E}}
 \fidelity{\phi}{\mathcal{E}(\psi)}\geq1-\Delta,
\end{eqnarray}
where $\Delta=1-\left(1-\epsilon-\sqrt{\delta}\right)^2$.
By using a lemma about the convex approximation \cite[Lemma 3]{SGS23-2} in the similar way to derive Eq.~\eqref{eq:cond2}, we can verify that Eq.~\eqref{eq:cond1+2} is equivalent to
\begin{equation}
\label{eq:cond1+2-1}
 \min_{J(\mathcal{E})\in\mathcal{E}_{E\rightarrow B}}\max_{\phi\in\puredop{\hh_B}}\min_{\rho\in\dop{\hh_E}}
 \lpnorm{tr}{\mathcal{E}(\rho)-\phi}\leq\Delta.
\end{equation}
Let $\mathcal{E}(\rho)=\sum_{\hat{\phi}\in E}\tr{M_{\hat{\phi}}\rho}\hat{\phi}$ maximize Eq.~\eqref{eq:cond1+2-1}.
Eq.~\eqref{eq:cond1+2-1} implies that
\begin{equation}
 \max_{\phi\in\puredop{\hh_B}}\min_p\lpnorm{tr}{\sum_{\hat{\phi}\in E}p(\hat{\phi})\hat{\phi}-\phi}\leq\Delta.
\end{equation}
Since this implies that the convex hull of $E$ forms a $\Delta$-net of $\puredop{\hh_B}$, the theorem about the optimal convex approximation \cite[Theorem 1]{SGS23-2} implies that $E$ is a $\sqrt{\Delta}$-net of $\puredop{\hh_B}$, i.e.,
\begin{equation}
  \max_{\phi\in\puredop{\hh_B}}\min_{\hat{\phi}\in E}\lpnorm{tr}{\hat{\phi}-\phi}\leq\sqrt{\Delta}.
\end{equation}
By using Lemma \ref{lemma:net}, we obtain
\begin{equation}
\log_2(Dd^2)^2=\log_2(\dim\hh_B\dim\hh_E)^2\geq\log_2|E|\geq(d-1)\log_2\left(\frac{1}{\Delta}\right).
\end{equation}
\end{proof}

\section{Discussion}
We investigated how the input dimension of a strong disentangler relates to the output dimension and approximation parameters. Firstly, we presented explicit constructions of the strong disentangler based on $\epsilon$-net and quantum de Finetti theorem. This shows the existence of a strong disentangler with an exponential input dimension compared to the output dimension. Secondly, we proved that any strong disentangler with certain approximation parameters must have such an exponential input dimension. We achieved this by reducing the strong disentangler into an approximately entanglement-breaking channel. This provides an important partial proof to the original disentangler conjecture since the class of strong disentanglers is a subset of that of disentanglers that is wide enough to contain all the known disentanglers.
As mentioned in the introduction, the disentangler is an approximately entanglement-annihilating channel, which differs from an approximately entanglement-breaking channel in general. Therefore, it is important to conduct further research on finding more connections between these two types of channels in order to resolve the disentangler conjecture.

\section{Acknowledgements}
We thank Yoshifumi Nakata, Takaya Matsuura, Mio Murao, Koji Azuma, Hayata Yamasaki, Tomoyuki Morimae, Ryuhei Mori, Takuya Ikuta, Yuki Takeuchi, and Yasuhiro Takahashi for their helpful discussions.
This work was partially supported by JST Moonshot R\&D MILLENNIA Program (Grant no.JPMJMS2061).
SA was partially supported by JST, PRESTO Grant no.JPMJPR2111 and MEXT Q-LEAP Grant no. JPMXS0120319794.
GK was supported in part by the Grant-in-Aid for Scientific Research (C) no.20K03779, (C) no.21K03388, and (S) no.18H05237 of JSPS, and CREST (Japan Science and Technology Agency) Grant no.JPMJCR1671.
ST was partially supported by JSPS KAKENHI Grant nos. JP20H05966 and JP22H00522.
\appendix

\section{Fidelity distance from the set of separable states}
We derive alternative formulation of the maximal fidelity to separable states, which is essentially the minimal Bures distance to separable states, as following.
\begin{lemma}
\label{lemma:fidelity}
 For given $\rho\in\dop{\hh_A\otimes\hh_B}$, by letting $\ketPhiS{ABE}$ be a purification of $\rho$, it holds that for any integer $r\geq(d_Ad_B)^2$,
 \begin{equation}
 \max_{\sigma\in\sep{A}{B}}F(\rho,\sigma)=\max_{\{\psiSI{B}{i},M_i\}_{i=1}^r}\sum_{i=1}^r\tr{(\psiSI{B}{i}\otimes M_i )\PhiS{BE}},
\end{equation}
where $\{M_i\in\pos{\hh_E}\}_{i=1}^r$ is POVM. Note that $\{M_i\}_i$ can be restricted as rank-1 POVM if $r\geq d_E$.
\end{lemma}
\begin{proof}
Owing to the Caratheodory's theorem, we can find the closest separable state to $\rho$ in the form of $\sigma=\sum_{i=1}^sp_i\phiSI{A}{i}\otimes\psiSI{B}{i}$ if $s\geq (d_Ad_B)^2$. Thus, by assuming $s\geq (d_Ad_B)^2$, the left hand side is equivalent to
 \begin{eqnarray}
&&\max_{\{p_i,\phiSI{A}{i},\psiSI{B}{i}\}_{i=1}^s}\fidelity{\rho}{\sum_{i=1}^sp_i\phiSI{A}{i}\otimes\psiSI{B}{i}}\\
&=&
\begin{cases}
 \max_{\{p_i,\phiSI{A}{i},\psiSI{B}{i}\}_{i=1}^s}\max_{V:\hh_E\rightarrow\hh_{E'}}\fidelity{\rho}{\sum_{i=1}^sp_i\phiSI{A}{i}\otimes\psiSI{B}{i}}\\
 a
\end{cases}\\
  &=&\max_{\{p_i,\phiSI{A}{i},\psiSI{B}{i},\tilde{\eta}_i\}_{i=1}^s}\left|\sum_i\sqrt{p_i}\braket{\Phi}{\phiSI{A}{i}}\ket{\psiSI{B}{i}}\ket{\tilde{\eta}_i}_E\right|^2,\\
  \label{eq:fidelity1}
  &=&\max_{\{p_i,\phiSI{A}{i},\psiSI{B}{i},\tilde{\eta}_i\}_{i=1}^s}\left(\sum_i\sqrt{p_i}\left|\braket{\Phi}{\phiSI{A}{i}}\ket{\psiSI{B}{i}}\ket{\tilde{\eta}_i}_E\right|\right)^2,
\end{eqnarray}
where $\{\tilde{\eta}_i\}_{i=1}^s$ is chosen from a set of orthonormal states if $s<d_E$ or a rank-1 POVM if $s\geq d_E$. By using Cauchy-Schwarz inequality,
\begin{eqnarray}
 \eqref{eq:fidelity1}&=&\max_{\{\phiSI{A}{i},\psiSI{B}{i},\tilde{\eta}_i\}_{i=1}^s}\sum_{i=1}^s\left|\braket{\Phi}{\phiSI{A}{i}}\ket{\psiSI{B}{i}}\ket{\tilde{\eta}_i}_E\right|^2\\
 \label{eq:fidelity2}
 &=&\max_{\{\psiSI{B}{i},\tilde{\eta}_i\}_{i=1}^s}\sum_{i=1}^s\tr{(\psiSI{B}{i}\otimes\tilde{\eta}_i)\Phi_{BE}}.
\end{eqnarray}
In both case ($s<d_E$ and $s\geq d_E$), Eq.\eqref{eq:fidelity2} is upper bounded by the right hand side of the lemma with $r=s$. This implies $(LHS)\leq (RHS).$ On the other hand, since general POVM $\{M_i\}_{i=1}^r$ can be decomposed into a rank-1 POVM and coarse graining, the right hand side of the lemma is upper bounded by Eq.\eqref{eq:fidelity2} with $s=rd_E$. This implies $(RHS)\leq (LHS)$.
\end{proof}


\bibliographystyle{unsrt} 
\bibliography{references.bib}

\begin{thebibliography}{10}

\bibitem{HHH96}
Micha\l\ Horodecki, Pawe\l\ Horodecki, and Ryszard Horodecki.
\newblock Separability of mixed states: necessary and sufficient conditions.
\newblock {\em Physics Letters A}, 223(1):1--8, 1996.

\bibitem{AS17}
Guillaume Aubrun and Stanis\l aw~Szarek.
\newblock Dvoretzky’s theorem and the complexity of entanglement detection.
\newblock {\em DISCRETE ANALYSIS}, 1:20, 2017.

\bibitem{G04}
Leonid Gurvits.
\newblock Classical complexity and quantum entanglement.
\newblock {\em Journal of Computer and System Sciences}, 69(3):448--484, 2004.
\newblock Special Issue on STOC 2003.

\bibitem{G10}
Sevag Gharibian.
\newblock Strong {NP}-hardness of the quantum separability problem.
\newblock {\em Quantum Info. Comput.}, 10(3):343--360, March 2010.

\bibitem{KMY01}
Hirotada Kobayashi, Keiji Matsumoto, and Tomoyuki Yamakami.
\newblock Quantum certificate verification: Single versus multiple quantum
  certificates.
\newblock {\em ArXiv}, quant-ph/0110006, 2001.

\bibitem{ABDFS08}
Scott {Aaronson}, Salman {Beigi}, Andrew {Drucker}, Bill {Fefferman}, and Peter
  {Shor}.
\newblock The power of unentanglement.
\newblock {\em Theory of Computing}, 5:1--42, 2009.

\bibitem{BT09}
Hugue {Blier} and Alain {Tapp}.
\newblock All languages in {NP} have very short quantum proofs.
\newblock In {\em 2009 Third International Conference on Quantum, Nano and
  Micro Technologies}, pages 34--37, 2009.

\bibitem{B08}
Salman Beigi.
\newblock {NP} vs {QMA}$_{\log}$(2).
\newblock {\em Quantum Info. Comput.}, 10(1):141--151, January 2010.

\bibitem{GNN12}
Fran\c{c}ois Le~Gall, Shota Nakagawa, and Harumichi Nishimura.
\newblock On {QMA} protocols with two short quantum proofs.
\newblock {\em Quantum Info. Comput.}, 12(7--8):589--600, July 2012.

\bibitem{MW05}
Chris Marriott and John Watrous.
\newblock Quantum {A}rthur---{M}erlin games.
\newblock {\em Comput. Complex.}, 14(2):122--152, June 2005.

\bibitem{MZ10}
Lenka Morav{\v{c}}{\'{\i}}kov{\'{a}} and M{\'{a}}rio Ziman.
\newblock Entanglement-annihilating and entanglement-breaking channels.
\newblock {\em Journal of Physics A: Mathematical and Theoretical},
  43(27):275306, jun 2010.

\bibitem{H99}
Alexander~S. {Holevo}.
\newblock Coding theorems for quantum communication channels.
\newblock In {\em Proceedings. 1998 IEEE International Symposium on Information
  Theory (Cat. No.98CH36252)}, pages 84--, 1998.

\bibitem{HSR03}
Michael Horodecki, Peter~W. Shor, and Mary~Beth Ruskai.
\newblock Entanglement breaking channels.
\newblock {\em Reviews in Mathematical Physics}, 15(06):629--641, 2003.

\bibitem{HNW18}
Aram~W. Harrow, Anand Natarajan, and Xiaodi Wu.
\newblock {Limitations of Semidefinite Programs for Separable States and
  Entangled Games}.
\newblock {\em Communications in Mathematical Physics}, 366(2):423--468, 2019.

\bibitem{CKMR07}
Matthias Christandl, Robert K{\"{o}}nig, Graeme Mitchison, and Renato Renner.
\newblock {One-and-a-Half Quantum de Finetti Theorems}.
\newblock {\em Communications in Mathematical Physics}, 273(2):473--498, 2007.

\bibitem{C10}
Giulio Chiribella.
\newblock On quantum estimation, quantum cloning and finite quantum de finetti
  theorems.
\newblock In Wim van Dam, Vivien~M. Kendon, and Simone Severini, editors, {\em
  Theory of Quantum Computation, Communication, and Cryptography}, pages 9--25,
  Berlin, Heidelberg, 2011. Springer Berlin Heidelberg.

\bibitem{CFS02}
Carlton~M. Caves, Christopher~A. Fuchs, and R\"{u}diger Schack.
\newblock Unknown quantum states: The quantum de {F}inetti representation.
\newblock {\em Journal of Mathematical Physics}, 43(9):4537--4559, 2002.

\bibitem{R07}
Renato Renner.
\newblock {Symmetry of large physical systems implies independence of
  subsystems}.
\newblock {\em Nature Physics}, 3(9):645--649, 2007.

\bibitem{BP10}
Fernando G. S.~L. Brand{\~{a}}o and Martin~B. Plenio.
\newblock {A Generalization of Quantum Stein's Lemma}.
\newblock {\em Communications in Mathematical Physics}, 295(3):791--828, 2010.

\bibitem{APF04}
Andrew~C. Doherty, Pablo~A. Parrilo, and Federico~M. Spedalieri.
\newblock Complete family of separability criteria.
\newblock {\em Phys. Rev. A}, 69:022308, Feb 2004.

\bibitem{SGS23-2}
Seiseki Akibue, Go~Kato, and Seiichiro Tani.
\newblock Probabilistic state synthesis based on optimal convex approximation,
  2023.

\end{thebibliography}

\end{document}